# Half Heusler Alloys for Efficient Thermoelectric Power Conversion

Long Chen,[1,a)] Xiaoyu Zeng,[2] Terry M. Tritt,[2,3] and S. Joseph Poon[1,a)]

[1] *Department of Physics, University of Virginia, Charlottesville, Virginia 22904-4714*

[2] *Department of Physics and Astronomy, Clemson University, Clemson, South Carolina 29634-0978*

[3] *Materials Science & Engineering Department, Clemson University, Clemson, South Carolina 29634*

Abstract

Half-Heusler (HH) phases (space group F43m, $Cl_b$) are increasingly gaining attention as promising thermoelectric materials in view of their thermal stability and environmental benignity as well as efficient power output. Until recently, the verifiable dimensionless figure of merit ($ZT$) of HH phases has remained moderate near 1, which limits the power conversion efficiency of these materials. We report herein ZT~1.3 in n-type (Hf,Zr)NiSn alloys near 850 K developed through elemental substitution and simultaneously embedment of nanoparticles in the HH matrix, obtained by annealing the samples close to their melting temperatures. Introduction of mass fluctuation and scattering centers play a key role in the high $ZT$ measured, as shown by the reduction of thermal conductivity and increase of thermopower. Based on computation, the power conversion efficiency of a n-p couple module based on the new n-type (Hf,Zr,Ti)NiSn particles-in-matrix composite and recently reported high-$ZT$ p-type HH phases is expected to reach 13%, comparable to that of state-of-the-art materials, but with the mentioned additional materials and environmental attributes. Since the high efficiency is obtained without tuning the microstructure of the Half-Heusler phases, it leaves room for further optimization.

Keywords: Half-Heusler, thermal conductivity, thermoelectric generators, energy conversion

---

[a)] Author to whom correspondence should be addressed;
Long Chen: lc4wn@virginia.edu; S. Joseph Poon: sjp9x@virginia.edu

The demand for energy produced from fossil fuels continues to contribute significantly to global carbon dioxide emissions. Thermoelectric conversion has gained much attention as one of the approaches for clean energy production by converting waste heat directly into electricity. RNiSn (R= Hf, Zr and Ti) Half-Heusler phases have attracted attention due to their thermal stability [1, 2], high reproducibility and potentially large power output [3-6]. The combination of large Seebeck coefficients with moderately low electrical resistivities and thermal conductivity in these materials has stimulated the quest for high dimensional figure of merit ($ZT$) [7-10]. $ZT$ is defined as $ZT=(S^2\sigma/\kappa)T$, where $S$, $\sigma$, and $\kappa$ are Seebeck coefficient, electrical conductivity, and thermal conductivity, respectively. $S^2\sigma$ (or $S^2/\rho$) is the power factor, where $\rho$ is electrical resistivity. Until recently, the highest verifiable $ZT$ of HH alloys has remained near 1 in the intermediate to high temperature range [4, 9, 10]. Despite the moderate $ZT$, Poon *et al.* demonstrated a maximum power conversion efficiency of 8.7% for a thermoelectric generator (TEG) that used Half-Heusler (HH) alloys as n- and p-legs [8]. Cook *et al.* reported 20% conversion efficiency for a 3-stage TEG that utilized HH alloys in the high temperature stage [5]. These reported TEG efficiencies are already higher than those of state-of-the-art TEGs [11-14]. The power density output of HH alloy based TEG was reported to exceed 3 W/cm$^2$ [3]. These results are encouraging since the measured efficiency is only within a few percent of the theoretical value.

Recent achievements of $ZT$~1.2 in some familiar n-type RNiSn phases [2, 15, 16] and $ZT$~1.5 in new p-type FeNbSb phases in the high temperature range of 600$^o$C to 800$^o$C [6] have given promise to the prospect of higher TEG efficiency. For the n-type alloys, Chen *et al.* obtained high $ZT$~1.2 in Hf$_{0.6}$Zr$_{0.4}$NiSn$_{0.995}$Sb$_{0.005}$ by compacting and annealing their samples near the melting point to improve the structural order, which enhanced the carrier mobility [2]. Schwall and Balke achieved $ZT$~1.2 in Ti$_{0.5}$Zr$_{0.25}$Hf$_{0.25}$NiSn$_{0.998}$Sb$_{0.002}$ through intrinsic phase separation that enhanced phonon scattering, reducing the thermal conductivity [15]. Subsequent to the latter work, Gurth *et al* reported similar high $ZT$ value in (Zr,Ti)NiSn phases [16]. By taking advantage of the heavy (flat) d band in FeNbSb HH phase, Fu *et al.* was able to substitute Nb with a high content of heavier element Hf that resulted in $ZT$~1.5 in p-type FeNb$_{0.88}$Hf$_{0.12}$Sb and FeNb$_{0.86}$Hf$_{0.14}$Sb [6]. The high content of heavier dopant helps to optimize the power factor as well as reduce thermal conductivity. In this paper, we report high $ZT$~1.3 in n-type (Hf,Zr)NiSn phases through substitution of Hf and Zr by Ti and addition of ZrO$_2$ nanoparticles. Reduction of thermal conductivity due to enhanced phonon scattering by mass fluctuation and nanoparticles is observed. In addition, ZrO$_2$ nanoparticles serve as potential barriers for carrier scattering that enhances the thermopower [4]. By combining the newly developed n-type HH composite and recently reported p-type $ZT$~1.5 FeNb$_{0.88}$Hf$_{0.12}$Sb in a TE (Thermoelectric) module, a power conversion efficiency near 13% is calculated with hot side temperature of about 670 $^o$C and cold side temperature of 40 $^o$C.

Ingots of Hf$_{0.6}$Zr$_{0.4}$NiSn$_{0.995}$Sb$_{0.005}$ and Hf$_{0.65}$Zr$_{0.25}$Ti$_{0.15}$NiSn$_{0.995}$Sb$_{0.005}$ were made by arc melting from the appropriate quantities of elemental Hf, Zr, Ti, Sn and pre-melted Sn$_{90}$Sb$_{10}$ alloy under argon atmosphere. A Sn-Sb alloy instead of elemental Sb was used in the alloying process in view of the relatively minute amount of Sb in the Half-Heusler alloys. Then, the ingots were pulverized into fine powders of around 10 to 50 µm in sizes. Both Hf$_{0.6}$Zr$_{0.4}$NiSn$_{0.995}$Sb$_{0.005}$ and Hf$_{0.65}$Zr$_{0.25}$Ti$_{0.15}$NiSn$_{0.995}$Sb$_{0.005}$ powders were mixed with 2 vol% ZrO$_2$ nano-particles. The powders with and without ZrO$_2$ nano-particles were then consolidated by using Spark Plasma Sintering (Thermal Technologies SPS 10-4) technique. These Hf$_{0.65}$Zr$_{0.25}$Ti$_{0.15}$NiSn$_{0.995}$Sb$_{0.005}$ samples were first sintered at a lower temperature of 1300 $^o$C for 15 min under 60 MPa to ensure that a single phase was formed in the mixed-phase ingots, followed by annealing at a higher temperature of 1350 $^o$C for another 15 min. On the other hand, the Hf$_{0.6}$Zr$_{0.4}$NiSn$_{0.995}$Sb$_{0.005}$ samples were known to have higher melting points and they were sintered directly at 1350 $^o$C for 30 min under 60 MPa. To confirm the phases of samples, an X-ray diffraction analysis was performed using the PANalyticalX'Pert Pro MPD (Multi Purpose Diffractometer) instrument in air at room temperature. The sample composition and microstructure was investigated using FEI Quanta 650 Scanning Electron Microscope. The electrical resistivity and thermoelectric power were measured by a four-probe method on ZEM3 system. The thermal conductivity was calculated from the specific heat $C_p$ (Netzsch Differential Scanning Calorimeter), the thermal diffusivity $\alpha$ (Netzsch LFA 457 MicroFlash system), and the sample density $\rho$ as $\kappa = C_p \alpha \rho$. The lattice thermal conductivity $\kappa_L = \kappa - \kappa_e$ can be achieved by knowing $\kappa_e$, where $\kappa_e$ is the electrical contribution that can be estimated by using the Wiedenann-Franz relationship $\kappa_e = L\sigma T$, where $L$ is the Lorenz Number. The unique Lorenz number for each of the samples was determined by using the equation proposed by Kim *et al.* [17].

The X-ray patterns of Hf$_{0.65}$Zr$_{0.25}$Ti$_{0.15}$NiSn$_{0.995}$Sb$_{0.005}$ with 0% and 2% ZrO$_2$ nanoparticle inclusions can be indexed to HH structure (space group F$\bar{4}$3m, C1$_b$). In Fig. 1, an impurity phase can also be identified as ZrO$_2$ in the n-type alloy dispersed with ZrO$_2$ nanoparticle inclusions. The compositional homogeneity checked by energy dispersive spectroscopy map scan over 10000 μm² area shows no evidence for any compositional variation. In addition, previous study has been done on the calculation of the binodal and spinodal curves in the isopleths TiNiSn-ZrNiSn and TiNiSn-HfNiSn [16]. Based on the calculation, the SPS temperature of 1350 °C for our $Hf_{0.65}Zr_{0.25}Ti_{0.15}NiSn_{0.995}Sb_{0.005}$ samples is sufficiently high above the calculated critical point in order to obtain single phase regarding to the Ti content. The typical main matrix in our samples has micron size grains, while for nanoclusions, they vary from 70 to 250 nm. Those grain sizes values are similar to those reported earlier [4].

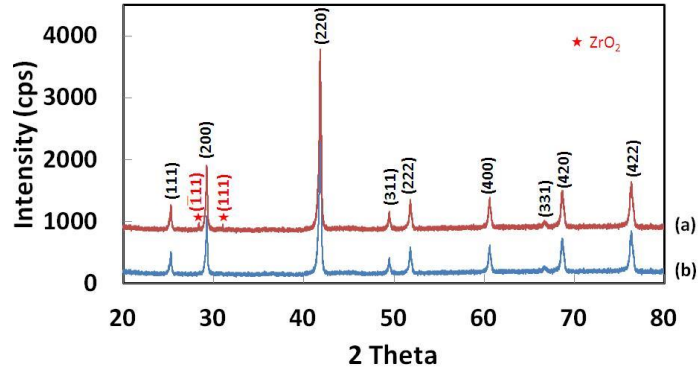

**Fig. 1** X-ray patterns of n-type $Hf_{0.65}Zr_{0.25}Ti_{0.15}NiSn_{0.995}Sb_{0.005}$ embedded with 2% (a) and 0% (b) $ZrO_2$ nanoparticles.

The temperature dependencies of the electrical resistivity $\rho$, the Seebeck coefficient $S$, and the power factor are shown in Fig. 2. It can be seen that all the $Hf_{0.6}Zr_xTi_{0.4-x}NiSn_{0.995}Sb_{0.005}$ compounds that contain or do not contain $ZrO_2$ nanoparticles show n-type behavior. As we replaced 15% Zr with Ti and added $ZrO_2$ nanoparticles, the resistivity increases over the entire temperature range. The Seebeck coefficient increases as the temperature rises from room temperature, reaches the maximum at about 780K, and then starts to decrease for each sample. The enhancement of the Seebeck coefficient by the addition of $ZrO_2$ nanoparticles serving as potential barriers for carrier scattering is also observed among these samples. The Hall coefficient ($R_H = -1/nq$) reveals both the carrier type and carrier concentration, where $n$ is the carrier concentration and $q$ is the carrier charge. The carrier mobility is then deduced from the relation ($\sigma = nq\mu_H$), where $\sigma$ is the electrical conductivity. The estimated errors for carrier concentration and mobility by Hall measurement are 10%. To estimate the effective band mass, the formulas below are used by assuming a single-band model with acoustic phonon scattering [18]:

$$S = \pm \frac{k_B}{e}\left[\frac{2F_1(\eta_F)}{F_0(\eta_F)} - \eta_F\right] \quad (1)$$

$$F_n(\eta_F) = \int_0^\infty \frac{x^n}{1+e^{(x-\eta_F)}} dx \quad (2)$$

$$n = \frac{4}{\pi}\left(\frac{2\pi m^* k_B T}{h^2}\right)^{\frac{3}{2}} F_{1/2}(\eta_F) \quad (3)$$

Where $F_n(\eta_F)$ is the Fermi-Dirac integral, $\eta_F$ is the reduced Fermi level defined as $\eta_F = E_F/k_B T$, $k_B$ is the Boltzmann constant, $m^*$ is the effective band mass. $h$ is the Planck constant, and $T$ is the temperature in K. The use of the single-band model for the above analysis was justified by previous study on the same system of materials [19]. The $n$, $\mu_H$, and calculated $m^*$ values are shown in Table I. The effective band mass for these alloys has been estimated to be larger than $2m_e$ [19, 20]. It shows that by Ti substitution and addition of ZrO$_2$, $n$ decreases and $\mu_H$ increases. The decrease of $n$ may be attributed to bandstructure effect in the case of alloying and charge trapping in the case of nanoparticles embedment. A plausible reason for the increase of carrier mobility could be due to the decrease in carrier scattering as n decreases. The thermal power can be written in the form of the scaling relation *S ~ qm\*T/n$^{2/3}$* after applying the Mott formula to a degenerate semimetal at temperature below the Fermi Temperature [21]. This relation shows that *S* increases as *n* decreases, as observed. Despite of increasing *S*, *PF=S$^2$/ρ* of the samples with Ti or ZrO$_2$ nanoparticles remain similar compared with Hf$_{0.6}$Zr$_{0.4}$NiSn$_{0.995}$Sb$_{0.005}$ samples due to the increase in $\rho$.

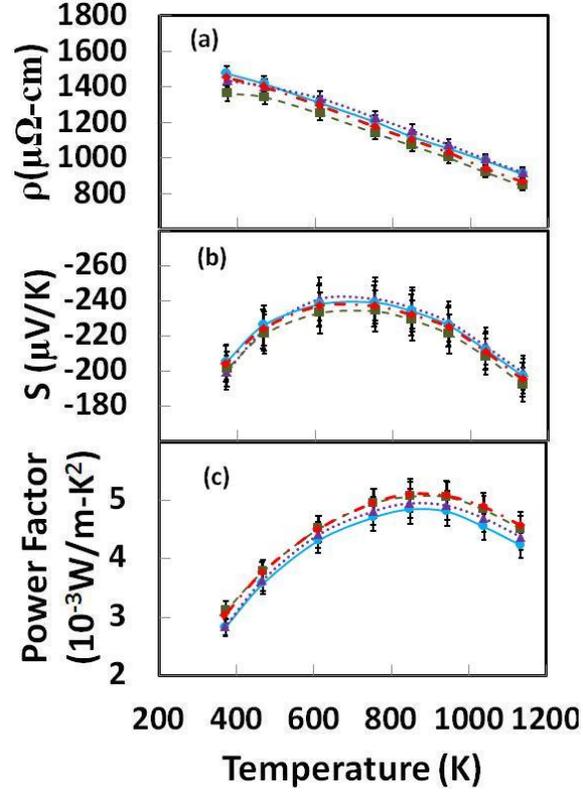

**Fig. 2** Thermoelectric properties of n-type $Hf_{0.6}Zr_{0.4}NiSn_{0.995}Sb_{0.005}$ (dashed line with green square), $Hf_{0.6}Zr_{0.4}NiSn_{0.995}Sb_{0.005}$/nano-$ZrO_2$ (dash dotted line with red rhombus), $Hf_{0.65}Zr_{0.25}Ti_{0.15}NiSn_{0.995}Sb_{0.005}$ (solid line with blue circle), $Hf_{0.65}Zr_{0.25}Ti_{0.15}NiSn_{0.995}Sb_{0.005}$/nano-$ZrO_2$ (round dotted line with purple triangle): (a) Electrical resistivity ($\rho$), (b) Seebeck coefficient ($S$), and (c) Power Factor.

Table I. Hall coefficient ($R_H$), carrier concentration ($n$), Hall mobility ($\mu_H$), and effective band mass ($m^*$) of $Hf_{0.6}Zr_{0.4}NiSn_{0.995}Sb_{0.005}$ and $Hf_{0.65}Zr_{0.25}Ti_{0.15}NiSn_{0.995}Sb_{0.005}$ embedded with 0% and 2% $ZrO_2$ samples.

| Composition | $R_H$ ($10^{-1}cm^3/C$) | $n$ ($10^{19}cm^{-3}$) | $\mu_H$ ($cm^2/(V*s)$) | $m^*$ ($m_0$) |
|---|---|---|---|---|
| $Hf_{0.6}Zr_{0.4}NiSn_{0.995}Sb_{0.005}$ | -5.17 | 1.21 ± 0.12 | 38.2 ± 3.8 | 2.53 |
| $Hf_{0.6}Zr_{0.4}NiSn_{0.995}Sb_{0.005}+2\%ZrO_2$ | -5.34 | 1.17 ± 0.12 | 38.4 ± 3.8 | 2.49 |
| $Hf_{0.6}Zr_{0.25}Ti_{0.15}NiSn_{0.995}Sb_{0.005}$ | -5.73 | 1.09 ± 0.11 | 38.9 ± 3.9 | 2.62 |
| $Hf_{0.6}Zr_{0.25}Ti_{0.15}NiSn_{0.995}Sb_{0.005}+2\%ZrO_2$ | -5.85 | 1.07 ± 0.11 | 40.7 ± 4.1 | 2.56 |

Thermal conductivity results are shown in Fig. 3(a). It is observed that the value of the thermal conductivity decreases with the substitution of Ti and addition of $ZrO_2$ nanoparticles in the compositions. The lattice thermal conductivity $\kappa_L$ was calculated from the relation $\kappa_L = \kappa - \kappa_e$ as shown in Fig. 3(b). It reveals that the lattice thermal conductivity decreases after substituting (Hf, Zr) with Ti and adding $ZrO_2$ nanoparticles in the compositions. The decrease of the phonon part of the thermal conductivity comes from the mass fluctuation scattering resulting from the substitution of (Hf, Zr) by Ti and the $ZrO_2$ nanoparticles serving as scattering centers.

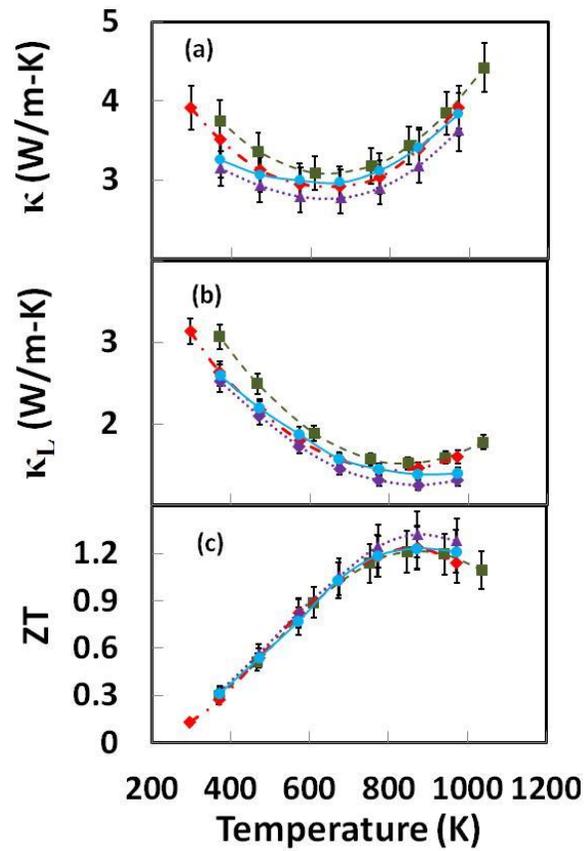

**Fig. 3** (a) Thermal conductivity ($\kappa$), (b) lattice thermal conductivity ($\kappa_L$), and (c) Figure of merit ZT of n-type $Hf_{0.6}Zr_{0.4}NiSn_{0.995}Sb_{0.005}$ (dashed line with green square), $Hf_{0.6}Zr_{0.4}NiSn_{0.995}Sb_{0.005}$/nano-$ZrO_2$ (dash dotted line with red rhombus), $Hf_{0.65}Zr_{0.25}Ti_{0.15}NiSn_{0.995}Sb_{0.005}$ (solid line with blue circle), $Hf_{0.65}Zr_{0.25}Ti_{0.15}NiSn_{0.995}Sb_{0.005}$/nano-$ZrO_2$ (round dotted line with purple triangle). Given the uncertainty in all the measurements, the resulting *ZT* has an uncertainty of $\approx \pm 10\%$, which is comparable or less than most other groups.

Due to the appreciable decrease of thermal conductivity, the figure of merit *ZT* increases despite the slight decrease of the power factor, as shown in Fig. 3(c). The maximum *ZT* values for $Hf_{0.6}Zr_{0.4}NiSn_{0.995}Sb_{0.005}$ and $Hf_{0.65}Zr_{0.25}Ti_{0.15}NiSn_{0.995}Sb_{0.005}$ are about 1.21 and 1.23 at 750K respectively and increase to 1.24 and 1.32 at 750K as a result of adding $ZrO_2$ nanoparticles. The measurement uncertainties are estimated to be 3% for $\rho$, $\alpha$, and $C_p$ and 5% for *S*. As a result, the resulting *ZT* contains 10% uncertainty, which is the same as that reported by Yan *et al.* [22].

Previously, device efficiency was calculated using a temperature-averaged figure of merit (*ZT*) [23, 24]. However, that is inadequate since all the relevant TE parameters are temperature dependent. Furthermore, the TE parameters for the n-leg and p-leg are also different. Here the device efficiency is calculated using the temperature dependent parameters for the n-leg and p-leg of a couple module. Starting with the following equations:

$$\text{Open circuit voltage: } V_{oc} = \int_{T_c}^{T_h}(S_p(T) - S_n(T))dT \quad (4)$$

$$\text{Current: } I = V_{oc} / (R_L + R_n + R_p) \quad (5)$$

$$\text{Output power: } P = I^2 R_L \quad (6)$$

The net heat absorbed can be estimated from the sum of the Peltier, Fourier, and Joule heat terms, thus heat input to the TE generator:

$$Q_{IN} = S_h I T_h + \bar{K} * \Delta T - 0.5 I^2 (R_n + R_p) \quad (7)$$

Where $S_h$ is the Seebeck coefficient at the hot-side temperature $T_h$.

$$\bar{K} = \frac{1}{(T_h - T_c)} \int_{T_c}^{T_h} \left( k_n(T) * \frac{A*\theta}{L} + k_p(T) * \frac{A}{L} \right) dT \quad (8)$$

$$R_n = \frac{1}{(T_h - T_c)} \int_{T_c}^{T_h} \left( \rho_n(T) * \frac{L}{A*\theta} \right) dT \quad (9)$$

$$R_p = \frac{1}{(T_h - T_c)} \int_{T_c}^{T_h} \left( \rho_p(T) * \frac{L}{A} \right) dT \quad (10)$$

$R_L$, $R_n$, and $R_p$ are the external resistance, electric resistance of n-type material, electric resistance of p-type material. $T_h$ and $T_c$ are the hot side and cold side temperature. $A$ is the cross-sectional area of p-type material. $\theta$ is the cross-sectional area ratio of n-type over p-type. $L$ is the length of each arm.

In this calculation, we assume one thermoelectric unicouple with one p- and one n-type leg which is shown in Fig. 4.

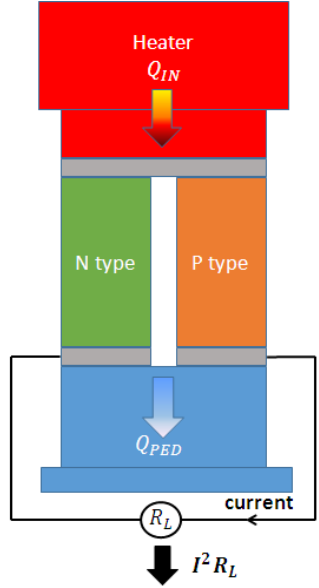

**Fig. 4** Schematic diagram of a one-stage TEG module.

The device performance can be written as:

$$\eta = \frac{output\ power}{heat\ input\ to\ the\ generator} \quad (11)$$

Apply (6) and (7) to the equation above, we can achieve:

$$\eta = \frac{I^2 R_L}{S_h I T_h + \bar{K} * \Delta T - 0.5 I^2 (R_n + R_p)} \quad (12)$$

The maximum efficiency can be obtained at the optimized external resistance and area ratio satisfying:

$$\frac{\partial \eta}{\partial R_L} = 0 \quad (13)$$

$$\frac{\partial \eta}{\partial \theta} = 0 \quad (14)$$

After solving the two equations above, we can achieve that when

$$\theta = (\overline{k_p} * \overline{\rho_n} / \overline{k_n} * \overline{\rho_p})^{1/2} \quad (15)$$

$$R_L = \left(1 + \Delta T \left(\frac{\bar{S}}{(\overline{\rho_n} * \overline{k_n})^{\frac{1}{2}} + (\overline{\rho_p} * \overline{k_p})^{\frac{1}{2}}}\right)^2 \left(\frac{S_h * T_h}{\bar{S} * \Delta T} - \frac{1}{2}\right)\right)^{1/2} (R_n + R_p) \quad (16)$$

The maximized performance can be expressed as:

$$\eta = \frac{\Delta T}{T_h} * \frac{\left(1 + \Delta T * \left(\frac{\bar{S}}{(\overline{\rho_n} * \overline{k_n})^{\frac{1}{2}} + (\overline{\rho_p} * \overline{k_p})^{\frac{1}{2}}}\right)^2 \left(\frac{S_h * T_h}{\bar{S} * \Delta T} - \frac{1}{2}\right)\right)^{\frac{1}{2}} - 1}{\frac{S_h * T_h}{\bar{S} * \Delta T}\left(\left(1 + \Delta T * \left(\frac{\bar{S}}{(\overline{\rho_n} * \overline{k_n})^{\frac{1}{2}} + (\overline{\rho_p} * \overline{k_p})^{\frac{1}{2}}}\right)^2 \left(\frac{S_h * T_h}{\bar{S} * \Delta T} - \frac{1}{2}\right)\right)^{\frac{1}{2}} + 1\right) - \frac{\Delta T}{T_h}} \quad (17)$$

Where $\Delta T$, $\bar{S}$, $\overline{k_n}$, $\overline{\rho_n}$, $\overline{k_p}$, and $\overline{\rho_p}$ are defined as the following:

$$\Delta T = T_h - T_C \quad (18)$$

$$\bar{S} = \frac{1}{\Delta T} \int_{T_c}^{T_h} \left(S_p(T) - S_n(T)\right) dT \quad (19)$$

$$\overline{k_n} = \frac{1}{(T_h - T_c)} \int_{T_c}^{T_h} (k_n(T)) dT \quad (20)$$

$$\overline{\rho_n} = \frac{1}{(T_h - T_c)} \int_{T_c}^{T_h} (\rho_n(T)) dT \quad (21)$$

$$\overline{k_p} = \frac{1}{(T_h - T_c)} \int_{T_c}^{T_h} (k_p(T)) dT \quad (22)$$

$$\overline{\rho_p} = \frac{1}{(T_h - T_c)} \int_{T_c}^{T_h} (\rho_p(T)) dT \quad (23)$$

If the n-type and p-type materials with the same corresponding *ZT* are selected, the cross-sectional area of n-type is found to be the same as that of p-type material. Then the expression for maximized device efficiency above will converge to that reported by Kim *et al* [25]. To compare the calculation with experiment, the efficiency of a p-n module consisting of p-type $Hf_{0.3}Zr_{0.7}CoSn_{0.3}Sb_{0.7}$/nano-$ZrO_2$ (*ZT*~0.8) and n-type $Hf_{0.6}Zr_{0.4}NiSn_{0.995}Sb_{0.005}$ (*ZT*~1) are calculated using the data from ref. 4. Results are shown in Fig. 5(a). It can be seen that result obtained from calculation shows a similar trend as the measurement data [4] but slightly higher. This difference is due to the perfect contact assumption in our model, thus results in a higher efficiency. The efficiency calculated by using the temperature-averaged figure of merit results in a slightly higher value compared to the value calculated by using equation (17). Even though the difference is small in this case, in the system with dramatic change of TE parameters over temperature, the deviation will be larger. We then applied our method to a p-n module that uses combination of our new n-type $Hf_{0.65}Zr_{0.25}Ti_{0.15}NiSn_{0.995}Sb_{0.005}$/nano-$ZrO_2$ and recently reported p-type $ZT$~1.5 $FeNb_{0.88}Hf_{0.12}Sb$ [6]. A power conversion efficiency 12.9% is achieved with hot side temperature of about 670 °C and cold side temperature of 40 °C as shown in Fig. 5(b)

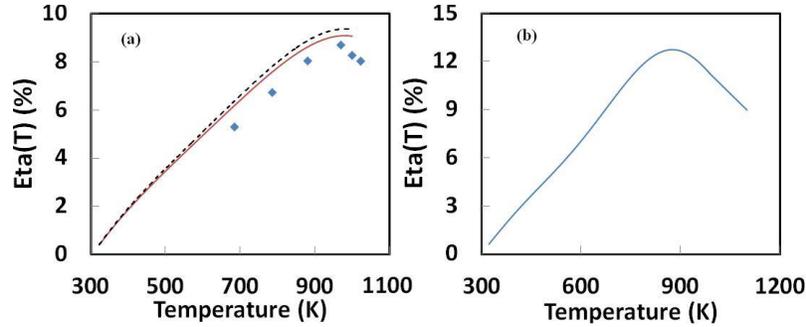

**Fig. 5** Heat-to-electric conversion efficiency for a single p-n couple made from (a) p-type $Hf_{0.3}Zr_{0.7}CoSn_{0.3}Sb_{0.7}$/nano-$ZrO_2$ and n-type $Hf_{0.6}Zr_{0.4}NiSn_{0.995}Sb_{0.005}$, experimental data (rhombus), calculation by using the temperature-averaged figure of merit (dashed line), calculation by using the temperature dependent parameters (solid line) and (b) p-type ZT~1.5 $FeNb_{0.88}Hf_{0.12}Sb$ and n-type $Hf_{0.65}Zr_{0.25}Ti_{0.15}NiSn_{0.995}Sb_{0.005}$/nano-$ZrO_2$an as a function of hot-side temperature.

In summary, thermoelectric properties of the substitution of Ti for (Hf, Zr) sites and the addition of nano-$ZrO_2$ in (Hf, Zr)NiSn have been measured. Due to enhanced phonon scattering by Ti substitution and embedded nanoparticles, the thermal conductivity is significantly reduced. In addition, $ZrO_2$ nanoparticles serve as potential barriers for carrier scattering that enhances the thermopower. A maximum $ZT$ of 1.3 at around 850K for $Hf_{0.65}Zr_{0.25}Ti_{0.15}NiSn_{0.995}Sb_{0.005}$/nano-$ZrO_2$ is obtained. The approach of annealing the materials near the melting point can be later applied to systems with less or no Hf to reduce the cost. A model of power conversion efficiency that takes into account the different temperature dependences of the thermoelectric parameters of the n-type and p-type materials is presented. Using the model to analyze a unicouple that uses the newly developed n-type and recently reported p-type HH materials yields a power conversion efficiency of 12.9% for hot-side temperature of about 890K is obtained. Given the agreement between calculation and experiment that involves similar alloys, the finding indicates an important step towards wider application of Half-Heusler materials for efficient power conversion devices. Further improvement is possible via tuning of the microstructure.